\documentclass[sigconf]{acmart}
\usepackage{color}
\usepackage{listings} 
\usepackage{graphicx}
\usepackage{subfigure}
\usepackage{booktabs}
\usepackage{todonotes}
\usepackage{amsfonts,amsthm}
\usepackage{textcomp}
\usepackage{xcolor}
\usepackage{multirow}
\usepackage{enumitem}
\usepackage{url}
\usepackage[ruled, linesnumbered]{algorithm2e}
\usepackage{bm}
\usepackage{diagbox}

\usepackage{soul} 

\AtBeginDocument{%
  \providecommand\BibTeX{{%
    \normalfont B\kern-0.5em{\scshape i\kern-0.25em b}\kern-0.8em\TeX}}}

\def\numx#1e#2{{#1}\mathrm{e}{#2}}

\setcopyright{rightsretained}
\acmJournal{PACMMOD}
\acmYear{2023} \acmVolume{1} \acmNumber{1} \acmArticle{6} \acmMonth{5} \acmPrice{}\acmDOI{10.1145/3588686}

\begin{document}

\title{Sequence-Based Target Coin Prediction for Cryptocurrency Pump-and-Dump}

\author{Sihao Hu}
\affiliation{%
  \institution{National University of Singapore}
  \country{}
}
\email{husihao26@gmail.com}

\author{Zhen Zhang}
\affiliation{%
  \institution{National University of Singapore}
  \country{}
}
\email{zhen@nus.edu.sg}

\author{Shengliang Lu}
\affiliation{%
  \institution{National University of Singapore}
  \country{}
}
\email{lusl@nus.edu.sg}

\author{Bingsheng He}
\affiliation{%
  \institution{National University of Singapore}
  \country{}
}
\email{dcsheb@nus.edu.sg}

\author{Zhao Li}
\affiliation{%
  \institution{Link2Do Technology}
  \institution{Zhejiang University}
  \country{}
}
\email{lzjoey@gmail.com}

\renewcommand{\shortauthors}{Hu et al.}

\begin{abstract}

With the proliferation of pump-and-dump schemes (P\&Ds) in the cryptocurrency market, it becomes imperative to detect such fraudulent activities in advance to alert potentially susceptible investors. In this paper, we focus on predicting the pump probability of all coins listed in the target exchange before a scheduled pump time, which we refer to as the \textit{target coin prediction} task. Firstly, we conduct a comprehensive study of the latest 709 P\&D events organized in Telegram from Jan. 2019 to Jan. 2022. Our empirical analysis reveals some interesting patterns of P\&Ds, such as that pumped coins exhibit intra-channel homogeneity and inter-channel heterogeneity. Here \textit{channel} refers a form of group in Telegram that is frequently used to coordinate P\&D events. 
This observation inspires us to develop a novel sequence-based neural network, dubbed SNN, which encodes a channel's P\&D event history into a sequence representation via the positional attention mechanism to enhance the prediction accuracy. Positional attention helps to extract useful information and alleviates noise, especially when the sequence length is long. Extensive experiments verify the effectiveness and generalizability of proposed methods. Additionally, we release the code and P\&D dataset on GitHub\footnote{\url{https://github.com/Bayi-Hu/Pump-and-Dump-Detection-on-Cryptocurrency}}, and regularly update the dataset.

\end{abstract}

\maketitle

\section{Introduction}

\begin{figure}[tbp]
\centering
\includegraphics[width=6.8cm]{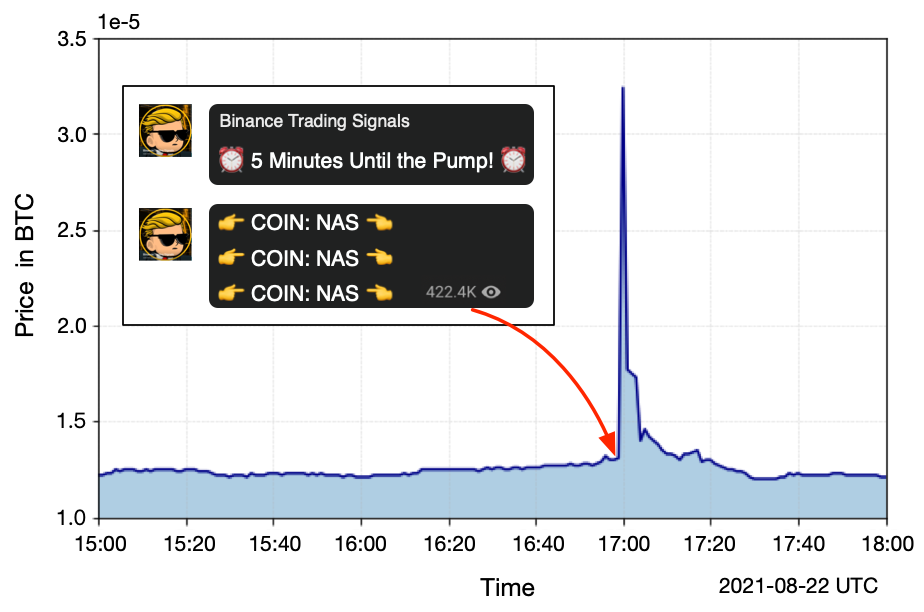}
\caption{The \$NAS pump-and-dump coordinated by the Telegram channel "Binance Trading Signals".}
\label{fig:PDcase}
\end{figure}

Pump-and-dump (P\&D) is a manipulative scheme that attempt to boost an asset price before selling the cheaply purchased assets at a higher price. 
While its origins can be traced back to the stock market~\cite{renault2014pump}, the burgeoning popularity of cryptocurrencies~\cite{ruan2020revealing,gupta2020building,attanasio2019quantitative} has led to the proliferation of P\&Ds within the cryptocurrency realm.
Organizing cryptocurrency P\&Ds typically took place in encrypted messaging platforms such as Telegram, where organizers create \textit{pump channels} to attract speculators. For a planned pump, its organizer posts an announcement several days in advance, clearly stating the target date, time, and exchange. As the scheduled time approaches, organizers use a \textit{countdown} strategy to intermittently remind members, without disclosing the target coin until the scheduled time. As illustrated in Figure~\ref{fig:PDcase}, a P\&D coordinated by the Telegram channel "Binance Trading Signals" (a channel with more than 300,000 subscribers) took place on Binance on August 22, 2021, at 17:00 UTC. After the coin NAS was disclosed, the price inflated immediately and reached 250\% in one minute, before dropping shortly afterward. Although the entire process lasted only five minutes, a staggering \$27 million was transferred between participants.

\begin{figure*}[tbp]
\centering
\includegraphics[width=17.7cm]{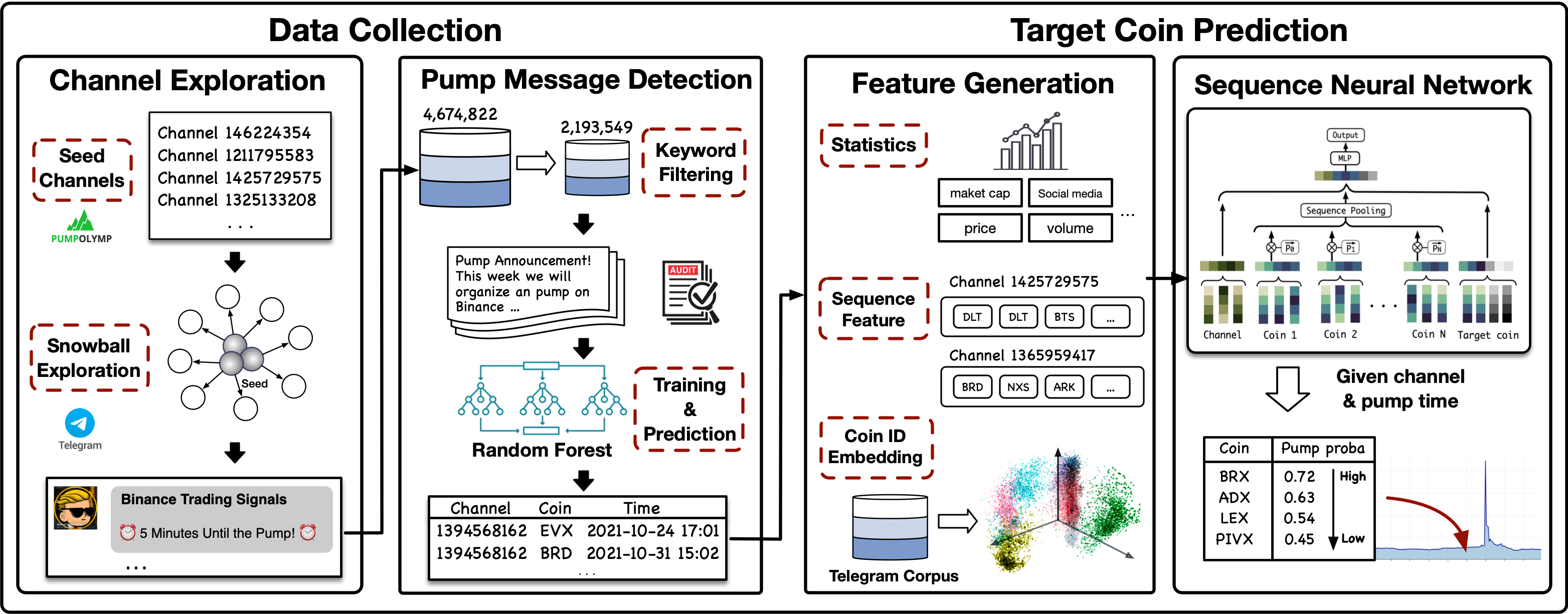}
\caption{The data science pipeline consists of two stages: data collection and target coin prediction.}
\vspace{-0.2cm}
\label{fig:workflow}
\end{figure*}

As P\&D schemes occur more frequently and attract greater public attention, it is imperative to detect such fraudulent activities~\cite{BERT4ETH,Turbo,ye2021gpu} and inform susceptible investors before they fall victim. Existing efforts~\cite{kamps2018moon,la2020pump,la2021doge,mirtaheri2021identifying} mainly focus on P\&D post-detection task,
which aims to detect whether a P\&D has happened or is happening.
However, we argue that this task fails to meet practical needs, as P\&Ds typically occur rapidly, leaving no time to alert investors.
In this paper, we concentrate on the \textit{target coin prediction} task, \textit{i.e.}, to predict the pump likelihood of all coins listed in the target exchange one hour before the pump time, given the exchange and pump time announced in advance.
The target coin prediction task is a interesting and challenging data science problem. To achieve accurate predictions, we rely on heterogeneous data sources, including historical statistics of trading price and volume, as well as text data that collected from Telegram. Different data sources require different tools and algorithms to extract useful and timely information. 


As shown in Figure~~\ref{fig:workflow}, we propose an efficient data science pipeline consisting of two main stages: data collection and target coin prediction. The data collection stage targets on identifying as many historical P\&Ds as possible. It explores pump channels based on verified seed channels, gathers tremendous text messages from Telegram, and identifies important pump messages through keyword filtering and machine learning algorithms. The data collection stage works offline, maintains a P\&D dataset, and updates it regularly. 

Given the pump messages identified in the data collection stage, the target coin prediction stage aims to make effective and efficient predictions. It collects multi-modal features from heterogeneous data sources, including historical trading prices and volumes from cryptocurrency market statistics, and coin embeddings pre-trained on the large text corpus collected from Telegram. We then generate features to feed into the prediction model and obtain the the pump probability of all coins listed on the target exchange one hour before the scheduled pump time. The entire process of target coin prediction can achieve real-time efficiency to ensure the timeliness.


Empirically, our data pipeline filters 709 P\&Ds out of 4,674,822 Telegram messages posted from Jan. 2019 to Jan. 2022. By analyzing them we discover that: 1) Coins with middle-cap value and high discussion degree in social media and forums are more likely to be targeted, indicating that the choice of the coin is well thought out; 2) The buy-in behavior of organizers starts within 60 hours before the scheduled pump time, and the pre-pump phenomenon confirms the existence of the hierarchy of Telegram channels. Precursors like the boost in price and trading volume caused by insiders (organizers and VIP members) can be a powerful signal for detecting P\&D coins in advance; 3) Coins pumped by the same channel exhibit homogeneity in both terms of statistics and semantics, whereas, coins selected by different channels exhibit distinct heterogeneity.


Prior to our work, P\&Ds are treated as isolated, unrelated events~\cite{xu2019anatomy,nghiem2021detecting}. However, our analysis show that a channel's pump coin selection follows some specific patterns and can be used to predict the future coin. Therefore, we construct channels' pump history as sequences and use sequential model to encode these sequences into embedding vectors to represent their coin selection strategies/preferences. We propose an effective Sequence Neural Network (SNN) with the core design of a positional attention module, which can directly capture the skip-correlation in one layer without serial computation as RNNs, and decouple the constraint of CNNs in the depth of layers and length of sequences. Secondly, we identify and address the \textit{coin-side cold start problem} that are largely overlooked by previous works but is essential in the practical settings. 



Experiments on a real-world P\&D dataset show that SNN outperforms its competitors with a large margin. To further verify the generalizability of the data science pipeline, we extend it to a different task, \textit{i.e.}, cryptocurrency price forecasting. Experiments show that our methodology is extensible and well-suited for other tasks that involves multiple sources of sequential data.

\noindent \textbf{Contributions:} To summarize, the contributions are as follows:
\begin{itemize}[leftmargin=8 pt]

\item \textbf{Pipeline.} We build a data science pipeline that extracts useful and timely information from heterogeneous data sources, to support effective and efficient target coin prediction.

\item \textbf{Analysis.} We present a comprehensive study on the latest P\&Ds, and uncover some interesting characteristics of P\&Ds such as intra-channel homogeneity and inter-channel heterogeneity.

\item \textbf{Model.} We propose SNN with a positional attention module to capture the pump history of channels when predicting the target coin. We identify and address the coin-side cold-start problem in practical settings.


\end{itemize}


\section{A Typical Pump-and-dump Process}
\label{sec:typical_PD}
\noindent \textbf{Set up:} A successful pump requires a large number of participants to buy the target cryptocoin at the same time.
Therefore, the organizer would establish a public channel, which is a specific form of group chat that only allows the organizer to post messages. 
Then the organizer recruits as many subscribers as possible by posting invitation links on social media like Twitter, or other channels for advertisement. 
Moreover, we notice that many pump organizers create another private VIP channels, which only accept paid members to subscribe.
In return, VIP members will be informed of the target coin name minutes or even hours before the coin name is released on the public channel.


\noindent \textbf{Pump Announcement:} When the public channel gets enough members, the pump organizer releases a pump announcement several days before a pump activity. The announcement usually includes three important information: the precise time to initiate the pump, the target exchange, and the pairing coin. As the scheduled pump time approaches, organizers keep reminding members of the pump event by counting down. The counting frequency gradually increases from once a day to once a few hours, and then once a few minutes before the pump time.
During this process, the organizers advise members to transfer sufficient pairing coin to the exchange and post some "pump rules" like "Do not sell immediately".



\noindent \textbf{Pre-pump:} VIP members in private channels know the target coin name in advance, so they can buy coins at a lower price and wait to sell them at a higher price at the pump time. If they buy a large volume of coins in a short time, it can lead to a price hike before the pump time, which we call "pre-pump."

\noindent \textbf{Pump:} Upon a few minutes before the pre-arranged pump time, the organizers will end the countdown with a typical message like: "The next message will be the coin name!"
The coin name is then posted at the pump time, usually in the uppercase symbol name of the coin, such as "FIC," and occasionally in the format of an OCR-proof image to prevent automatic recognition. Then the coin's price immediately surges, often reaching several times the original price within one minute.

\noindent \textbf{Dump:} Although the organizer may urge members to hold onto the coin and not sell it immediately, participants typically know that the price will quickly drop and do not want to sell later than others. Thus, just a few moments after the price reaches its peak, it starts to drop, causing more participants to panic-sell in turn. Eventually, the coin price will bounce back to the original value or even lower after the P\&D.



\section{Data Collection}

\subsection{Channel Exploration}

We explore the pump channels in Telegram, which is currently the largest platform for P\&D schemes to thrive~\cite{nizzoli2020charting,xu2019anatomy}. Firstly, we collect a total of 1,142 channels from PumpOlymp~\cite{pumpolymp}, a website that provides verified P\&D channels in Telegram. Then we use the Telethon API~\cite{telethon} to check their status. We find that nearly half of channels (564/1,142) have been deleted due to inactivity for over six months. Secondly, we adopt the \textit{snowball} strategy~\cite{nizzoli2020charting} to explore more pump channels, based on an observation that pump channel organizers sometimes post invitation links across other channels to attract more participants. Specifically, we collect all historical messages posted in seed channels and extract the invitation links from these seed pump channels to explore new channels. To ensure high relatedness, we set the exploration hop to 2. After filtering out the expired links, we get 137 channels that are still in active and not duplicated with the seed channels.
We retrieve a total of 4,674,822 messages that are posted on these 715 channels between Jan. 1, 2019, and Jan. 13, 2022.


\begin{table}[tbp]
\centering
\caption{Performance on Pump Message Detection}
\begin{tabular}{c|cccc}
\toprule
\textbf{Model} & \textbf{AUC} & \textbf{Precision} &\textbf{Recall} & \textbf{F1}  \\ 
\midrule
LR & 0.988 & 0.892 & 0.913 & 0.902   \\
RF & \textbf{0.994} & \textbf{0.901} & \textbf{0.939} & \textbf{0.920}   \\

\bottomrule 
\end{tabular}
\label{tab:pump_detection_result}
\end{table}
\begin{table}[tbp]
\centering
\caption{Statics of P\&D Dataset}
\begin{tabular}{ccccc}
\toprule
\textbf{\# Sample} & \textbf{\# Event} &\textbf{\# Channel} & \textbf{\# Coin} & \textbf{\# Exchange}  \\ 
\midrule
1,335 & 709 & 108 & 278 & 18  \\
\bottomrule 
\end{tabular}
\label{tab:dataset_statistic}
\end{table}

\begin{table}[tbp]
\small
\centering
\caption{Examples of P\&D Dataset}
\begin{tabular}{ccccc}
\toprule
\textbf{Channel} & \textbf{Coin} &\textbf{Exchange} & \textbf{Pair} & \textbf{Timestamp}  \\ 
\midrule

1394568162 & EVX & Binance & BTC & 2021-10-24 17:01:59 \\
1394568162 & BRD & Binance & BTC & 2021-10-31 15:01:30 \\
1394568162 & MTH & Binance & BTC & 2022-01-13 17:00:19 \\
... & ... & ... & ... & ... \\
1250076602 & EVX & Binance & BTC & 2021-10-24 17:00:07 \\
1250076602 & EZ & Binance & BTC & 2022-01-05 19:00:05 \\

\bottomrule 
\end{tabular}
\label{tab:dataset_example}
\vspace{0.1cm}
\end{table}

\subsection{Pump Message Detection}

The amount of Telegram messages is so tremendous that manual annotation becomes prohibitively expensive and time-consuming. 
Fortunately, the majority of pump-relevant messages conform to specific patterns that can be readily discerned by humans, rendering them identifiable by machine learning algorithms.

The workflow for pump message detection is illustrated in Figure~\ref{fig:workflow}: Firstly, a keyword matching strategy is employed to reduce the proportion of non-pump messages in the text dataset, which reserves any message that mentions a coin or exchange name, or includes keywords such as "pump," "target," "hold," "sell," etc. This simple but useful strategy selects 2,193,549 messages out of 4,674,822 messages.
Secondly, we randomly select and label 5,050 messages as ether pump and non-pump messages. Pump messages are labeled as those including pump announcements, pump countdowns, target coin releases, and post-pump reviews, while all other messages are labeled as non-pump messages.
Subsequently, we remove punctuation marks, stop words, URLs, and emojis from labeled messages and represent each message as a vector using the TF-IDF technique. We separate 70\% samples to train a Random Forest (RF) model and a Logistic Regression (LR) model, and test them on the 30\% left samples. 

The results are presented in Table~\ref{tab:pump_detection_result}, where metrics are calculated using a relatively low threshold of 0.2 to identify as many pump messages as possible. 
Because a single pump message is insufficient to identify a P\&D event, we aggregate messages from into \textit{session}s if the time interval between two adjacent messages is less than 24 hours. In this case, a session is the minimum time unit for a channel to hold a P\&D, and we assume that a P\&D occurs at most once in a session (which is verified by us manually). Finally, we identify 1,335 P\&D samples from total 2,006 sessions based on the 37,525 pump messages identified by the RF model. Each P\&D sample is represented by a quintuple (channel\_id, target, exchange, pairing coin, timestamp), as exemplified in Table~\ref{tab:dataset_example}. A summary of the dataset statistics is presented in Table~\ref{tab:dataset_statistic}.

\section{Pump-and-Dump Schemes Analysis}

This section presents analytical studies at the coin, event, and channel levels, aimed at answering the following research questions:

\begin{itemize}[leftmargin=8pt]
    \item \textbf{Q1}: What types of coins are susceptible to P\&D schemes?
    \item \textbf{Q2}: Is it possible to detect P\&D schemes in advance?
    \item \textbf{Q3}: Do the target coin selection strategies differ across channels?
\end{itemize}

\subsection{Coin-Level Analysis (Q1)}


\begin{figure}[tbp]
\centering
\includegraphics[width=8.4cm]{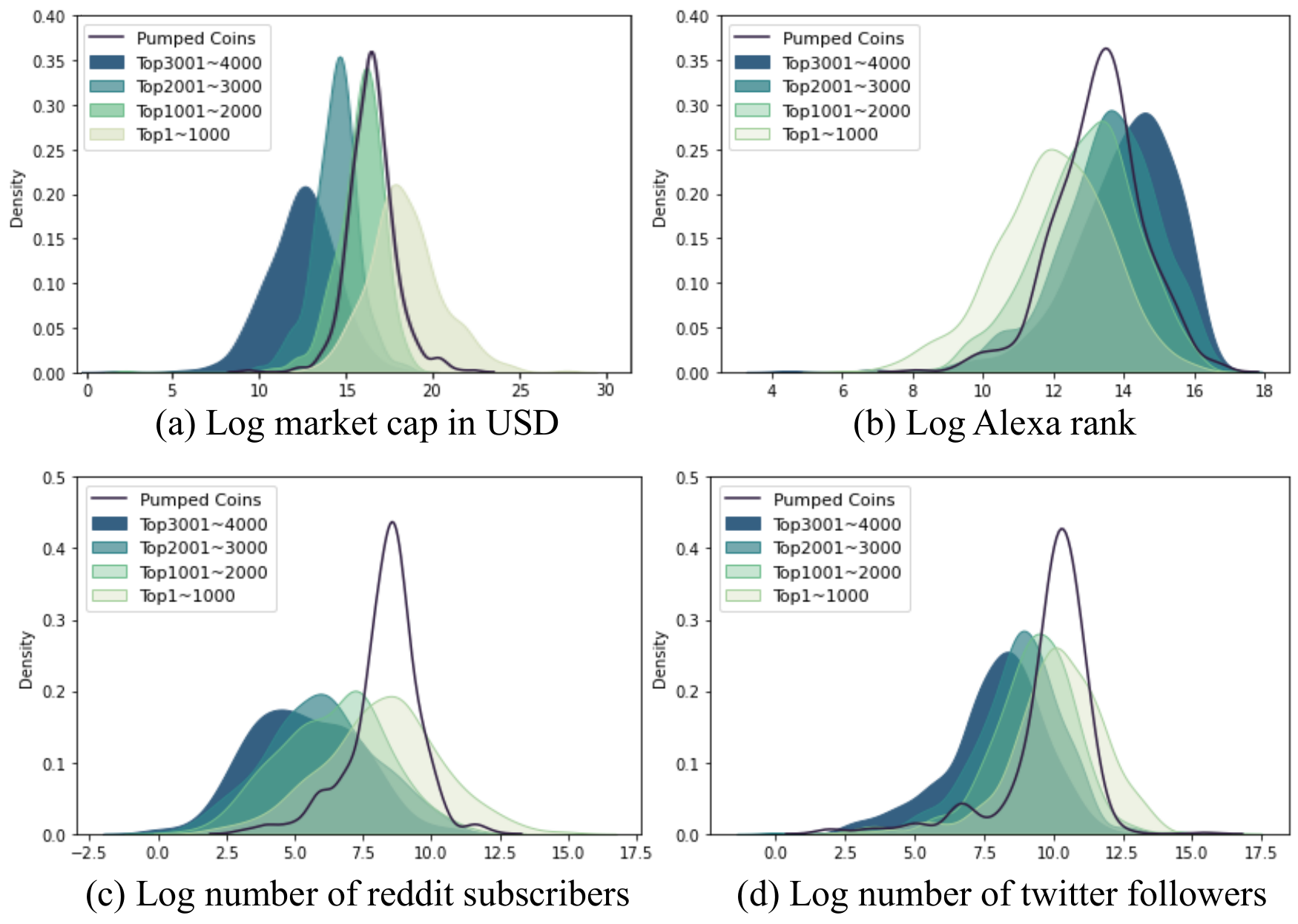}
\vspace{0.1cm}
\caption{Distributions of pumped coins and top-4000 coins on four features.}
\label{fig:coin_analysis}
\vspace{0.2cm}
\end{figure}

To investigate Q1, we utilize the CoinGecko~\cite{coingecko}
API to collect historical daily statistics on coins, such as market capitalization, volume data, and social media indices, etc. For pumped coins, we collect the data from three days preceding the pump time, as this data is generally more stable before P\&Ds. For a fair comparison, we randomly retrieve historical data from Jan. 1, 2019, to Jan. 13, 2021 for the top 4,000 coins ranked on the CoinGecko website (almost all pumped coins are ranked above 4000).
Figure~\ref{fig:coin_analysis} demonstrates the distribution of pumped coins and the top 4,000 coins based on several features. Due to space limitations, we only report on four features: market capitalization, Alexa rank (an index representing global web popularity), the number of Reddit subscribers, and the number of Twitter followers.

From Figure~\ref{fig:coin_analysis}(a) and (b), we observe that the market capitalization and Alexa rank distributions of pumped coins are most similar to those of coins ranked in the top1001-2000 range, suggesting that P\&D organizers tend to target middle-cap coins. This is because it is much more challenging to manipulate the prices of large-cap coins~\cite{li2021cryptocurrency,xu2019anatomy}, especially for pump channels with a small number of participants. Additionally, organizers rarely select coins with low market capitalization and web popularity (Alexa rank), which typically reflect low trading liquidity. For example, if the market is frozen, there will be no other traders being attracted to get involved and transactions will be limited to the group, making common members hesitant to engage in the pump due to the likelihood of losing money. Another observation, illustrated in Figure~\ref{fig:coin_analysis}(c) and (d), is that the distributions of numbers of Reddit subscribers and Twitter follower are similar to those of coins ranked in the top1-1000 range with minor variances, indicating that pumped coins have a significant presence on social media platforms. 
According to our statistics, the likelihood of a coin being pumped again is 60.1\%, indicating that even with data collect three days prior to the pump time, the coin might have already undergone a previous pump. Once a coin is pumped, it is widely discussed on social media, and pump participants spread misinformation about the coin to lure potential victims.

\begin{figure}[tbp]
\centering
\includegraphics[width=8.7cm]{./Figure/EventAnalysis.pdf}
\caption{The observational study of P\&D events.}
\label{fig:event_analysis}
\vspace{0.2cm}
\end{figure} 

\subsection{Event-Level Analysis (Q2)}

\textbf{Event Distribution Drift across Exchanges}: Out of the 709 P\&D events in our dataset, \textbf{62.8}\% (445) took place in Binance, 20.6\% (146) in Yobit, 8.7\%(62) in Hotbit and 3.0\%(21) in Kucoin, which differs significantly from previous literature: Xu et al.~\cite{xu2019anatomy} report 51\% of P\&Ds took place in Cryptopia, 27\% in Yobit, and only \textbf{17}\% in Binance from October 2018 to February 2019; Li et al.~\cite{li2021cryptocurrency} report 52.6\% of P\&Ds happened in Bittrex, 32.2\% in Yobit, and only \textbf{15.2}\% in Binance from May 2017 to August 2018; 
It is noteworthy that the proportion of P\&Ds occurring on Binance has drastically increased (from 15.2\% to 62.8\%) in the past few years. We identify two key factors that lead to this drift: firstly, Bittrex's ban on P\&Ds in Nov. 2017~\cite{Bittrex}, and secondly, Cryptopia's liquidation following a severe hack in Jan. 2019~\cite{coindesk2018cryptopia}. Additionally, Binance's advantages in terms of coin number, user volume, and commissions have made it the favored exchange for pump organizers.

The drift has impacted on the pattern of P\&D attacks. Previous studies show that attacks on Binance typically have the lowest average return rate, approximately 29\% of that on Yobit~\cite{xu2019anatomy,li2021cryptocurrency}, and involved an average of 1.42 channels per attack. This is likely due to the fact that Binance's trading volume is much larger than that of other exchanges, necessitating a greater number of participants for a successful P\&D. However, our analysis shows that each P\&Ds on Binance involves, on average, 2.25 channels, suggesting that pump organizers improve return rates by coordinating multiple channels for a single pump event.

\noindent \textbf{Precursors to P\&Ds:} After the organizer decides which coin to pump, they gradually purchase the coin and release the name of the coin in the private channel, allowing VIP members to buy in before the scheduled pump time. The purchase behavior of organizers and VIP members can cause significant market movements for middle-cap coins. To study the precursors to P\&Ds, we plot the time series of prices and trading volumes 72 hours before and 24 hours after the pump time in Figure~\ref{fig:event_analysis}(a) and (b). The time series are averaged from 445 pump events that occurred on Binance and were paired with BTC.
For each P\&D, we select the earliest announce time across multiple channels as the pump time if the event is organized by multiple channels, and retrieve historical OHLCV (open, high, low, close, volume) data for each minute using the Binance API~\cite{binancekline}.

One prominent finding is the gradual increase in coin price tens of hours prior to the pump time, as observed in Figure~\ref{fig:event_analysis}(a). In Figure~\ref{fig:event_analysis}(b), we note that the start point of frequent trading is about 57 hours before the pump time, with trading volume remaining relatively low before this point. Correspondingly, the average coin price in Figure~\ref{fig:event_analysis}(a) gradually starts to rise. We also notice that several hikes appear from 48 hours to 1 hour before the pump time, very similar to the largest trading volume hike around the pump time. We posit that these hikes are caused by the pre-pump events in which VIP members purchase a large number of coins in a short time. Figure~\ref{fig:event_analysis}(d) gives a verified pre-pump in a VIP channel 5 hours before the pump time. The observed boost in price and trading volume can be utilized as a powerful signal to detect P\&Ds in advance. To demonstrate this, we calculate the average returns within a time window from $x+1$ hours to $1$ hour before the pump time ($x=1,3,6,12,24,36,48,60,72$). We compare the results with randomly selected coins and timestamps, calculating the statistics in the same manner. As presented in Figure~\ref{fig:event_analysis}(c), the highest average return reaches 9.5\% when $x=60$, and the average returns of randomly selected coins are close to zero. 

\begin{figure}[tbp]
\centering
\includegraphics[width=8.7cm]{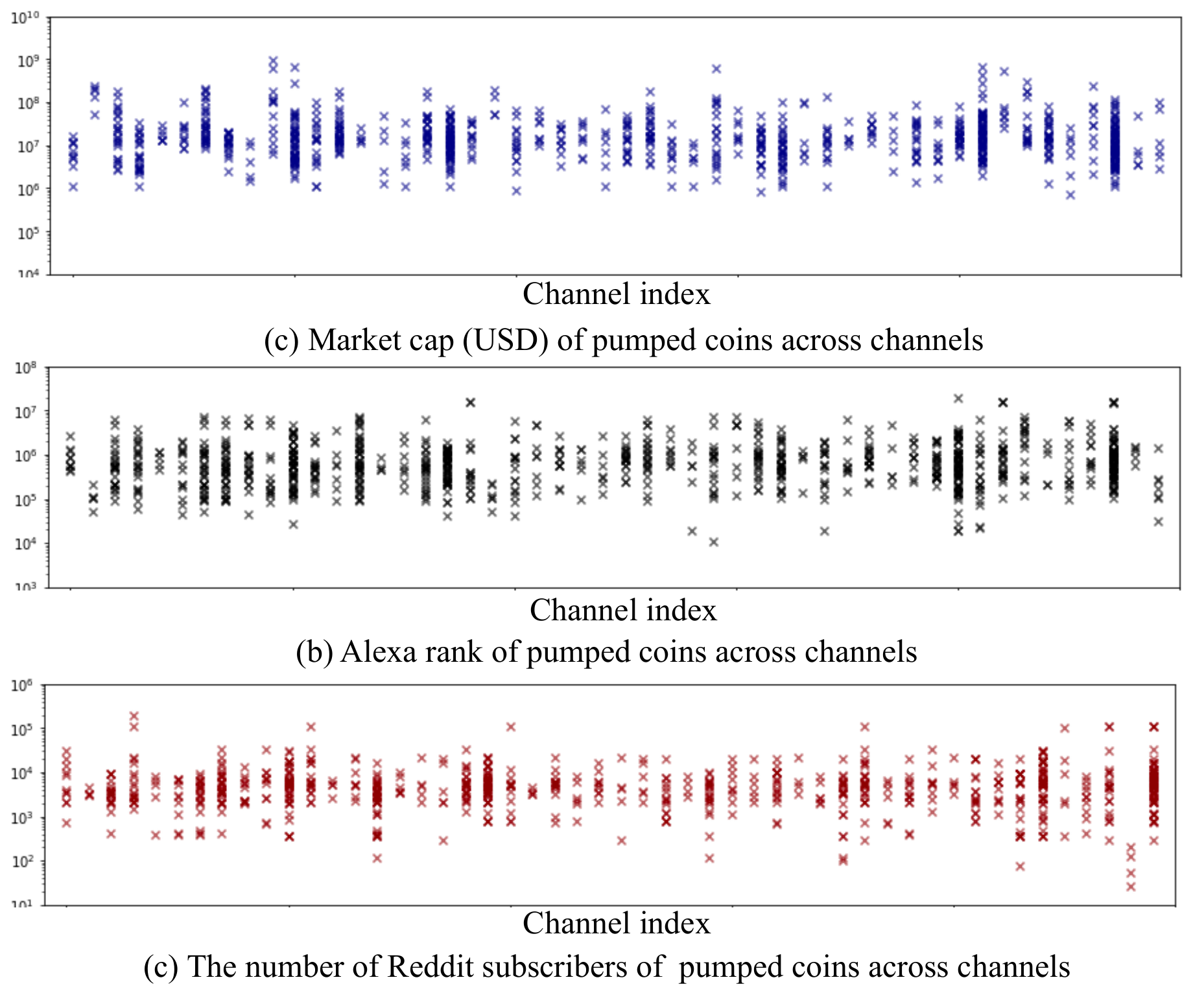}
\caption{Scatter plots of three features of pumped coins across different channels.}
\vspace{0.2cm}
\label{fig:ChannelAnalysis}
\end{figure} 

\begin{figure}[tbp]
\centering
\includegraphics[width=7.0cm]{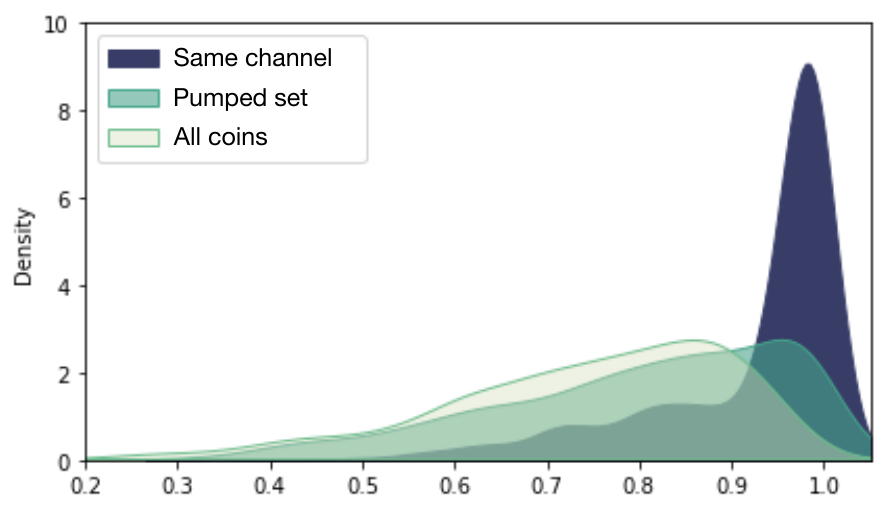}
\vspace{0.1cm}
\caption{Semantic similarity distributions of coin pairs selected under three strategies.}
\vspace{0.3cm}
\label{fig:semantic_similarity}
\end{figure} 


\subsection{Channel-Level Analysis (Q3)} To answer Q3, we investigate coin selection strategies of different channels in terms of several characteristics. Figure~\ref{fig:ChannelAnalysis}(a) presents a scatter plot of the market capitalization of the pumped coins in different channels. Each dot in the figure represents a pumped coin, with the channel index on the X-axis and the market capitalization value on the Y-axis. The first observation is that the market capitalization of coins pumped by a specific channel exhibits homogeneity (similarity) within a specific range, while exhibits heterogeneity across different channels. The underlying reason for this phenomenon is straightforward: the success of a pump event heavily relies on the number of participants. A channel with larger population of members generally has access to a wider selection of coins to pump and tends to target coins with larger market capitalization to generate higher trading volume, which is difficult feat for smaller channels to achieve.
Figure~\ref{fig:ChannelAnalysis}(b) and (c) show the Alexa rank and number of Reddit subscribers of the pumped coins respectively, which exhibit similar effects as Figure~\ref{fig:ChannelAnalysis}(a).


\noindent \textbf{Semantic Similarity:} We further investigate whether coins pumped by the same channel tend to share similar semantics. Specifically, we pre-train SkipGram embeddings~\cite{SkipGram} on a huge cryptocurrency-related corpus and extract the word embedding of the coin symbol to represent its semantics. Figure~\ref{fig:semantic_similarity} illustrates the cosine similarity distributions of pairs selected under three distinct strategies: 1) selection from the same channel's pump history; 2) selection from all the pumped coins; 3) random selection from all available coins. Evidently, the semantic similarity distribution of coins pumped by the same channel exhibit a significant high average value (0.92) with minor variance compared to pairs selected with strategy 2 (with average value of 0.80) and strategy 3 (with average value of 0.72). This empirical study further proves that a channel has its distinct coin selection pattern.

{\bf Summary:} We summarize the key findings:
\begin{itemize}[leftmargin=8pt]
    \item \textbf{A1}: Coins with middle capitalization and high levels of social media discussion are more likely to be targeted.
    \item \textbf{A2}: P\&Ds are predictable as the purchase behavior of organizers and VIP members can lead to significant market movement before the scheduled pump time.
    \item \textbf{A3}: Each pump channel has its distinct coin selection pattern that varies from other channels. The coins pumped by a given channel exhibit homogeneity in both statistics and semantics.
\end{itemize}

\section{Target Coin Prediction}

Inspired by our analysis, we in this section present a sequence-based neural network to leverage channels' historical coin selection information to predict the target coin. We also identify and address the coin-side cold-start problem to further improve the performance.

\subsection{Feature Generation}

The features can be categorized into three groups: channel, target coin, and sequence. Regarding channel features, we collect \textit{channel\_id} and \textit{subscriber\_num}; As for the target coin, we collect \textit{coin\_id} and stable statistics (\textit{market\_cap}, \textit{price}, \textit{volume}, \textit{Alex\_rank}, \textit{Twitter\_follower}, \textit{Reddit\_subscriber}, \textit{etc.}) three days prior to the pump event. We further calculate market movement signals (\textit{price}, \textit{return}, \textit{volume}, \textit{trade\_cnt}, \textit{etc.}) within ($x$+1) hour to 1 hour prior to the pump time with varing time window ranges ($x = [1,3,6,12,24,48,60,72]$).
For sequence features, we group pumped coins by \textit{channel\_id} and sort them chronologically. Features like \textit{coin\_id} and stable statistics are incorporated for pumped coins in the sequence.



\subsection{Sequence Neural Network (SNN)}
\label{sec:snn}
In SNN, we develop a sequence encoder that can extract a channel's historical pump sequence as a embedding vector to represent its coin selection pattern. As shown in Figure~\ref{fig:snn_model}, SNN mainly consists of several parts:

\begin{figure}[tbp]
\centering
\includegraphics[width=8.5cm]{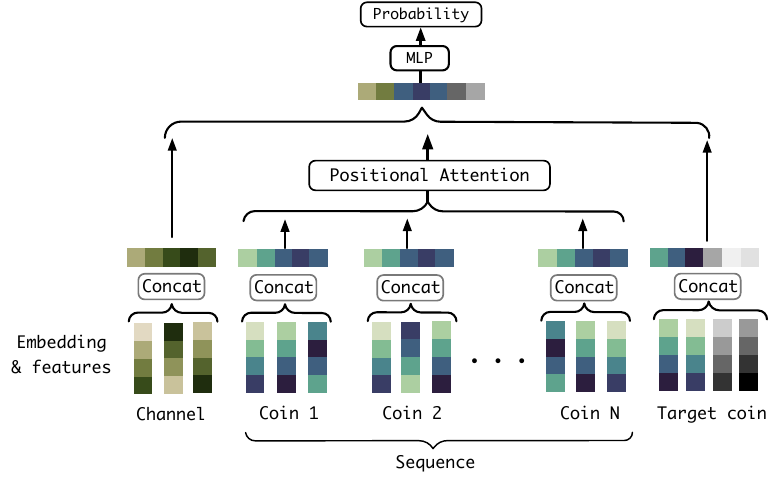}
\caption{Model architecture of SNN}
\vspace{0.2cm}
\label{fig:snn_model}
\end{figure}

\noindent \textbf{Embedding Layer:} The input feature contains categorical features, \textit{e.g.}, channel id and coin id, which cannot be directly used due to their high dimensionality. We adopt embedding techniques to embed the sparse features into low-dimensional dense vectors, which significantly eases computing. 
To reduce the redundancy of parameters, we make the sequence \textit{coin\_id} and target \textit{coin\_id} share the same latent space.
The generated embeddings are concatenated with other numeric features together to generate the overall features for the channel, target coin as follows:
\begin{equation}
    \boldsymbol{h}_{c} = h_{c}^{1} \oplus h_{c}^{2} \oplus ...\oplus h_{c}^{X}
\end{equation}
\begin{equation}
    \boldsymbol{h}_{t} = h_{t}^{1} \oplus h_{t}^{2} \oplus ...\oplus h_{t}^{Y}
\end{equation}
where $h_{c}^{i}$ and $h_{t}^{i}$ are the $i$-th feature attached to $\boldsymbol{h}_{c}$ and $\boldsymbol{h}_{t}$, respectively. $\oplus$ is the concatenation operator, $X$ and $Y$ are the number of features of channel and target coin. 
For coins in the sequence, we generate the whole sequence representation $\boldsymbol{h}_{s}$ by the positional attention module.

\begin{figure}[tbp]
\centering
\includegraphics[width=6.6cm]{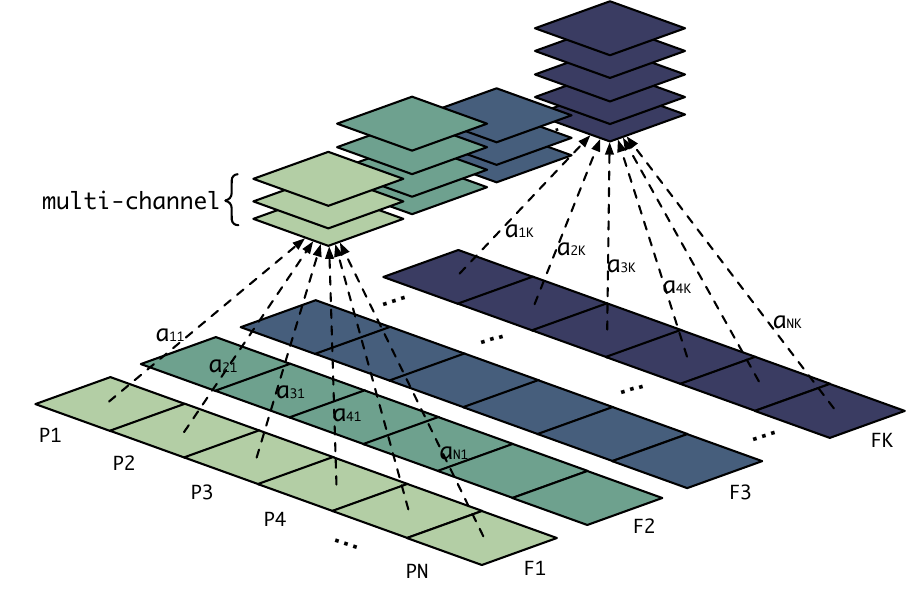}
\caption{Positional attention module}
\label{fig:pos_attention}
\end{figure} 

\noindent \textbf{Positional Attention:} A coin in a sequence has several features that may exhibit different sequential patterns. For example, we find that a channel might pump a specific coin periodically but never pump the coin continuously because otherwise it will be easily guessed by others. However, the organizer is likely to choose another coin with similar market capitalization for the next pump event, because the number of participants in the group remains stable over a period of time. This suggests the existence of two distinct patterns within the sequence: 1) temporal proximity pattern, suggesting that feature values that are closer in time are more likely to be correlated; 2) skip-correlated pattern, in which the most related timestamp is not the closest one but the previous ones, representing the periodicity or temporal delay.

Modeling skip-correlation is nontrivial in our scenario as this pattern widely exists in features collected from multiple data sources. Nevertheless, we find existing RNN-based methods are not well suited because the serial computation of RNNs makes them hard to retrieve previous information, especially when the sequence length and span of skip correlation are both large; Moreover,  CNN-based models (TCN~\cite{tcn}, WaveNet~\cite{wavenet}) require relative deep layers to cover the whole sequence as the sequence length goes longer. During this process, convolution operations across different features over and over again may hurt the expression of the pattern of a specific feature.

To capture skip-correlation and distinguish sequential patterns for different features to prevent interference, we propose a simple but effective encoder named \textit{positional attention}.
As presented in Figure~\ref{fig:pos_attention}, we utilize $P_1$, $P_2$, ..., $P_N$ to designate the $1, 2, ..., N$-th position of entities (coins) in the sequence, with each entity having a total of $K$ features denoted as $F_1$, $F_2$, ..., $F_K$. For ease of explanation, we use $F_{i,j}$ to indicate the $j$-th feature of $P_i$. For each feature $F_j$, we generate its positional attention vector $\boldsymbol{\alpha}_{j}$ as follows:
\begin{equation}
    \boldsymbol{a}_{j} = \left(\boldsymbol{f}(a_{1,j},a_{2,j}, ...,a_{N,j})\right)
    \label{eq:initialization}
\end{equation}
\begin{equation}
    \boldsymbol{\alpha}_{j} = \mathrm{Softmax}(\boldsymbol{a}_{j})
    \label{eq:softmax}
\end{equation}
where $a_{1,j}, a_{2,j}, ..., a_{N,j}$ are zero-initialized learnable parameters, and $\boldsymbol{f}(\cdot)$ is an adjustable mapping function ($\mathbb{R}^{N}\xrightarrow{f} \mathbb{R}^{N}$) such as an MLP layer. Since the $\mathrm{Softmax}(\cdot)$ is computed across different positions, the attention vector can take positional information into account. Subsequently, we multiply each element $\alpha_{i,j}$ in the attention vector with its corresponding $P_i$'s j-th feature $F_{i,j}$ as shown below:
\begin{equation}
\begin{aligned}
    h_j &= \boldsymbol{\alpha}_{j} \cdot (F_{1,j}, F_{2,j}, ..., F_{N,j})^{T}\\
     & = \sum_{i=1}^{N} \alpha_{i,j} \cdot F_{i,j} 
\end{aligned}
\end{equation}

As illustrated in Figure~\ref{fig:pos_attention}, for each feature $F_{j}$, $j\ \in \{1,2,...,K\}$, we initialize multiple attention vectors and repeat the attention computation multiple times to generate $C_{j}$ \textit{multi-channel} of $h_j$, denoted as $\boldsymbol{h}_{j}=(h_j^1, h_j^2, ..., h_j^{Cj})$. This approach can increase the capacity of the positional attention, as some of the features may exhibit both of the two aforementioned patterns. Unlike the number of channels for filters in CNN-based models, the choice of channel number in our positional attention is independent for different features. In practice, we find that this hyper-parameter can be set larger for those non-skip-correlated features, so that besides capturing its major pattern, some of the channels can also learn the minor, skip-correlated patterns. Finally, we flatten and concatenate the mutli-channel vectors of all the features to generate the final representation vector $\boldsymbol{h}_{s}$ for the entire sequence:
\begin{equation}
    \boldsymbol{h}_{s}= \boldsymbol{h}_1 \oplus ... \oplus \boldsymbol{h}_j \oplus... \oplus \boldsymbol{h}_K 
\end{equation}
where $\boldsymbol{h}_{j}$ is the multi-channel representation vector of $F_j$.

\noindent \textbf{MLP Layer:} The embedding and attention layers are primarily based on linear projections. We utilize several fully connected layers and activation functions to endow the model with non-linearity. The output $\hat{y}$ represents the predicted pump probability:
\begin{equation}
    \hat{y} = \operatorname{Sigmoid} \left( \operatorname{MLP}\left(\boldsymbol{h}_{c} \oplus \boldsymbol{h}_{t} \oplus \boldsymbol{h}_{s}\right) \right)
\end{equation}

\noindent \textbf{Loss Function:} The objective function used in SNN is the negative log-likelihood function defined as: 
\begin{equation}
L=-\frac{1}{|\mathcal{D}|} \sum_{(\hat{y}, y) \in \mathcal{D}}(y \log \hat{y}+(1-y) \log (1-\hat{y}))
\end{equation}
where $\mathcal{D}$ is the training set, and $y\in\{1,0\}$ is the ground-truth label that denotes whether it is a P\&D or not.

{\bf Summary:} we summarize the key advantages of SNN:
\begin{itemize}[leftmargin=8pt]
    \item \textbf{D1}: SNN can capture skip-correlation directly in one layer without the need for serial computation as RNNs and decouples the depth of layers from the length of sequences covered by CNNs.
    \item \textbf{D2}: SNN distinguishes the sequence patterns at the feature-level, reducing interference between different features.
    
    \item \textbf{D3}: The time complexity of SNN is $\mathcal{O}(N\cdot K\cdot C), (N \gg K,C)$, and the attentions for $K$ features and $C$ channels can be calculated in parallel, making it computationally efficient.
\end{itemize}


\subsection{Coin-Side Cold-Start Problem}

To ensure consistency with real-world applications, we evaluate our method on a testing set consisting of samples all collected after the training set. This strategy also prevents any future information leakage, since the sequences of some latter samples may include label information of former samples.
Under such practical settings, we observe the occurrence of the coin-side cold-start problem during end-to-end training. Similar to the item-side cold start problem in the recommendation field~\cite{hu2022gift}, the coin-side cold-start problem occurs in two cases: 1) coins that are pumped in the testing set are never be pumped in the training set; 2) coins in the testing set are never exist in the training set. Coins falling into either of these categories can cause the model to struggle in classifying them correctly due to the lack of robust representations to represent the coin appropriately. 

\begin{figure}[tbp]
\centering
\includegraphics[width=8.5cm]{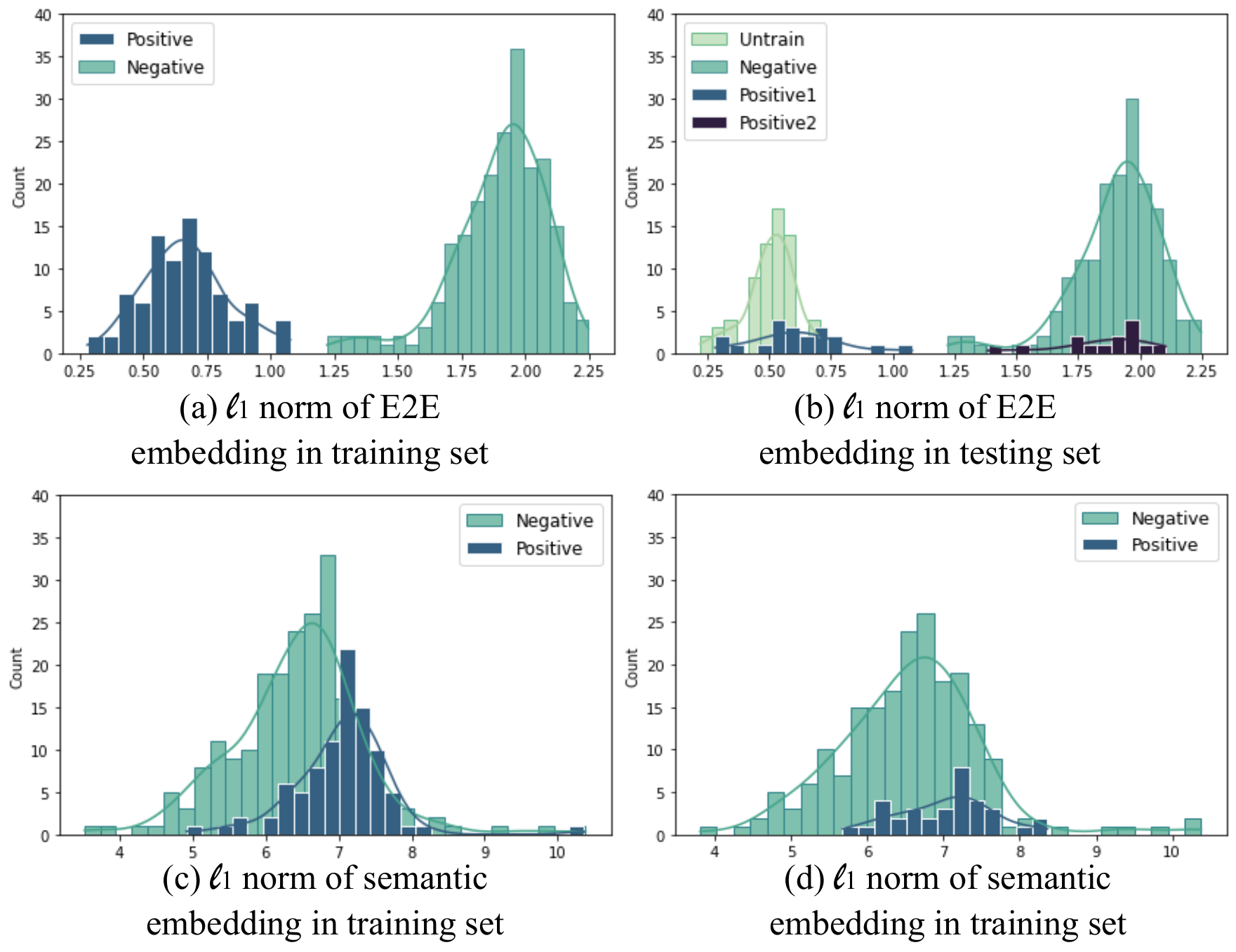}
\vspace{0.1cm}
\caption{$\ell_{1}$ norm distributions of two types of coin id embedding in training and testing sets.}
\vspace{0.2cm}
\label{fig:l1norm_visualization}
\end{figure}

Figure~\ref{fig:l1norm_visualization}(a) shows the $\ell_1$ norm of coin\_id embedding of positive (pumped) and negative coins in the training set, revealing a pronounced difference between two distributions; 
Figure~\ref{fig:l1norm_visualization}(b) illustrates the $\ell_1$ norm of coin\_id embeddings of four types of coins in the testing set. \textit{Untrain} denotes the coins never exist in the training set, thus their coin\_id embeddings remaining the same as random initialization. \textit{Positive1} represents positive coins that are pumped in the training set, while \textit{Positive2} represents the distribution of positive coins that are never pumped in the training set, which exhibit a similar distribution to that of the negative coins. The distributions of the \textit{untrain} and \textit{positive2} coins correspond to the two cases mentioned before and are the primary factors of the cold-start problem.

To address the cold-start problem, we propose to use the word embedding of a coin symbol to replace its coin\_id embedding learned during end-to-end (E2E) training.
Word embeddings learned on a large corpus can cover almost all of the coin symbols, making embedding vectors sufficiently trained and containing semantic information. Specifically, we utilize well-known word embedding techniques, \textit{e.g.}, SkipGram~\cite{SkipGram}, CBoW~\cite{CBOW}, to pre-train the embeddings on a large corpus of Telegram messages we collected from the cryptocurrency-related channels and groups, not limited to the pump channels.
Figure~\ref{fig:l1norm_visualization}(c) and Figure~\ref{fig:l1norm_visualization}(d) shows the $\ell_1$ norm distributions of SkipGram embeddings of positive and negative coins in the training set and testing set, respectively. Apparently, two distributions are consistent in the two sets, indicating that using word embedding can alleviate the coin-side cold-start problem.

\begin{table}[tbp]
\centering
\caption{P\&D Dataset Description}
\begin{tabular}{llllll}
\toprule
\textbf{Pumped?}& \textbf{Train} & \textbf{Validate} & \textbf{Test} & \textbf{Total}  & \textbf{Ratio} \\ 
\midrule
TRUE    & 648       & 100         & 200     & 948    & 0.48\%      \\
FALSE   & 106,900   & 24,666     & 64,099 & 195,665 & 99.52\%     \\ 
\midrule
Total   & 107,548   & 24,766     & 64,299 & 196,613 & 100.0\%    \\ 
\bottomrule
\end{tabular}
\vspace{0.4cm}
\label{tab:dataset_description}
\end{table}
\section{Experiments}

\subsection{Experiment Settings} 

\textbf{Dataset:} We limit our experiments on predicting coins that were pumped on Binance and paired with BTC. Following this setting, 948 samples are selected (71\%) out of the total 1,335 samples as the positive samples.
For each positive sample, we label other eligible coins listed on Binance at the time of pump events as negative coins and generate features for them to construct the negative samples. 
We split the training, validation, and testing sets by the timestamp "2021-01-19 00:00:00" and "2021-05-10 00:00:00".
Table~\ref{tab:dataset_description} presents the statistics of the dataset used in the experiment. 
The difference between positive and negative ratios across three sets is due to the varying numbers of coins listed on Binance during different periods.

\noindent \textbf{Competitors:} We compare the proposed SNN with two types of baselines. The first type is machine-learning methods using handcrafted features, such as Logistic Regression (LR) and Random Forest (RF). We use the mean encoding technique to compensate for the lack of embedding layers in these models. The second type of methods are deep-learning algorithms including a DNN using the same features as SNN except the sequence, RNN-based models such as LSTM, BiLSTM, GRU and BiGRU, and CNN-based method TCN~\cite{tcn}. For RNN models, the hidden dimension of cells is set to 32; For TCN, the depth of convolution layer is set to 3 with 16 channels per layer, the kernel size is set to 4 to cover a 20-length sequence. For SNN, the number of channel is set to 8.

\noindent \textbf{Evaluation Metrics:} We employ the Hit Ratio (HR@$k$) to measure the model performance. Specifically, we rank a positive sample and its corresponding negative samples as a list according to pump probabilities predicted by the model and use HR@$k$ to indicate whether the ground truth coin is included in the top $k$ samples. The reported HR@$k$ is averaged across all the lists with $k=1,5,10,20,30$. HR@$k$ is well suited to evaluate a model in a real-world setting.

\begin{table}[tbp]
\small
\centering
\caption{Performance Comparison for Target Coin Prediction}
\setlength{\tabcolsep}{0.85mm}
\begin{tabular}{lp{0.15cm}p{0.15cm}p{0.15cm}p{0.15cm}p{0.1cm}p{0.15}p{0.15}p{0.15}p{0.15}p{0.15}p{0.15}p{0.15}}
\toprule
{\textbf{Metric}}& \multicolumn{1}{|c}{\textbf{LR}} & \multicolumn{1}{c}{\textbf{RF}} & \multicolumn{1}{c}{\textbf{DNN}} & \multicolumn{1}{c}{\textbf{LSTM}} & \multicolumn{1}{c}{\textbf{BLSTM}} & \multicolumn{1}{c}{\textbf{GRU}}& \multicolumn{1}{c}{\textbf{BGRU}} &
\multicolumn{1}{c}{\textbf{TCN}}& \multicolumn{1}{c}{\textbf{SNN}} \\ 
\midrule


\multicolumn{1}{c|}{HR@1} & \multicolumn{1}{c}{0.156} &  \multicolumn{1}{c}{0.189} & \multicolumn{1}{c}{0.225} & \multicolumn{1}{c}{0.207} & \multicolumn{1}{c}{0.203} & \multicolumn{1}{c}{0.229} & \multicolumn{1}{c}{0.163} & 
\multicolumn{1}{c}{0.256} & \multicolumn{1}{c}{\textbf{0.260}} \\

\multicolumn{1}{c|}{HR@3} & \multicolumn{1}{c}{0.269} &  \multicolumn{1}{c}{0.348} & \multicolumn{1}{c}{0.278} & \multicolumn{1}{c}{0.339} & \multicolumn{1}{c}{0.344} & \multicolumn{1}{c}{0.339} & \multicolumn{1}{c}{0.335} & 
\multicolumn{1}{c}{0.348} & \multicolumn{1}{c}{\textbf{0.383}} \\

\multicolumn{1}{c|}{HR@5} & \multicolumn{1}{c}{0.322} &  \multicolumn{1}{c}{0.417} & \multicolumn{1}{c}{0.383} & \multicolumn{1}{c}{0.423} & \multicolumn{1}{c}{0.396} & \multicolumn{1}{c}{0.414} & \multicolumn{1}{c}{0.401} & 
\multicolumn{1}{c}{0.427} & \multicolumn{1}{c}{\textbf{0.465}} \\

\multicolumn{1}{c|}{HR@10} & \multicolumn{1}{c}{0.449} &  \multicolumn{1}{c}{0.537} & \multicolumn{1}{c}{0.498} & \multicolumn{1}{c}{0.551}  & \multicolumn{1}{c}{0.546} & \multicolumn{1}{c}{0.529} & \multicolumn{1}{c}{0.555}  & 
\multicolumn{1}{c}{0.573} & \multicolumn{1}{c}{\textbf{0.596}} \\ 
                            
\multicolumn{1}{c|}{HR@20} & \multicolumn{1}{c}{0.608} &  \multicolumn{1}{c}{0.687} & \multicolumn{1}{c}{0.626} & \multicolumn{1}{c}{0.648}  & \multicolumn{1}{c}{0.630} & \multicolumn{1}{c}{0.626} & \multicolumn{1}{c}{0.678}  & 
\multicolumn{1}{c}{0.692} & \multicolumn{1}{c}{\textbf{0.727}} \\

\multicolumn{1}{c|}{HR@30} & \multicolumn{1}{c}{0.662} &  \multicolumn{1}{c}{0.731} & \multicolumn{1}{c}{0.727} & \multicolumn{1}{c}{0.696}  & \multicolumn{1}{c}{0.696} & \multicolumn{1}{c}{0.714} & \multicolumn{1}{c}{0.709}  & 
\multicolumn{1}{c}{0.770} & \multicolumn{1}{c}{\textbf{0.797}} \\ 
\bottomrule
\label{tab:comparison_tcp}
\end{tabular}
\end{table}

\begin{table}[tbp]
\small
\centering
\caption{Coin Embedding Test on Testing Set}
\setlength{\tabcolsep}{1.8mm}
\begin{tabular}{lp{0.20cm}p{0.20cm}p{0.20cm}p{0.20cm}p{0.20cm}p{0.20}p{0.20}p{0.20}}  
\toprule
{\textbf{Metric}}& \multicolumn{1}{|c}{\textbf{E2E}} & \multicolumn{1}{c}{\textbf{CBOW}} & \multicolumn{1}{c}{\textbf{SG}} & \multicolumn{1}{c}{\textbf{SNN}} & \multicolumn{1}{c}{\textbf{SNN$_{C}$}} & \multicolumn{1}{c}{\textbf{SNN$_{S}$}}\\ 
\midrule


\multicolumn{1}{c|}{HR@1} & \multicolumn{1}{c}{0.0} & \multicolumn{1}{c}{0.035} & \multicolumn{1}{c}{0.043}  & \multicolumn{1}{c}{\underline{0.260}} & \multicolumn{1}{c}{0.256} & \multicolumn{1}{c}{\textbf{0.277}} \\

\multicolumn{1}{c|}{HR@3} & \multicolumn{1}{c}{0.0} & \multicolumn{1}{c}{0.090} & \multicolumn{1}{c}{0.115}  & \multicolumn{1}{c}{\underline{0.383}} & \multicolumn{1}{c}{0.391} & \multicolumn{1}{c}{\textbf{0.414}} \\

\multicolumn{1}{c|}{HR@5} & \multicolumn{1}{c}{0.013} & \multicolumn{1}{c}{0.133} & \multicolumn{1}{c}{0.176}  & \multicolumn{1}{c}{\underline{0.465}} & \multicolumn{1}{c}{0.499} & \multicolumn{1}{c}{\textbf{0.513}} \\ 

\multicolumn{1}{c|}{HR@10}& \multicolumn{1}{c}{0.057} & \multicolumn{1}{c}{0.253} & \multicolumn{1}{c}{0.286}  & \multicolumn{1}{c}{\underline{0.596}} & \multicolumn{1}{c}{0.617} & \multicolumn{1}{c}{\textbf{0.623}} \\
                            
\multicolumn{1}{c|}{HR@20}& \multicolumn{1}{c}{0.101} & \multicolumn{1}{c}{0.362} & \multicolumn{1}{c}{0.376}  & \multicolumn{1}{c}{\underline{0.727}} & \multicolumn{1}{c}{0.731} & \multicolumn{1}{c}{\textbf{0.739}} \\

\multicolumn{1}{c|}{HR@30}& \multicolumn{1}{c}{0.242} & \multicolumn{1}{c}{0.472} & \multicolumn{1}{c}{0.487}  & \multicolumn{1}{c}{\underline{0.797}} & \multicolumn{1}{c}{0.806} & \multicolumn{1}{c}{\textbf{0.823}} \\ 
\bottomrule
\label{tab:q3}
\end{tabular}
\end{table}

\begin{figure*}[tbp]
\centering
\includegraphics[width=18.0cm]{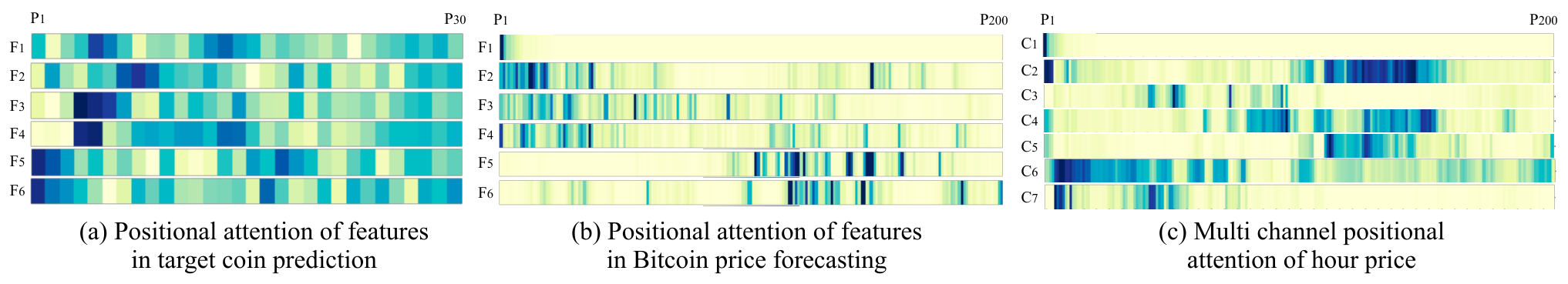}
\caption{Positional attention patterns of different features}
\label{fig:visualization}
\end{figure*}



\subsection{Performance Comparison}

Table~\ref{tab:comparison_tcp} summarizes the results of all the competitors on the testing set. From the table, we can observe that: 1) Among sequential models, RNN-based models (LSTM, BLSTM, GRU and BiGRU) perform worse than TCN and SNN, since RNN-based models fail to capture the skip-correlated pattern due to their serial computation paradigm; 2) SNN works better than TCN and other competitors by a margin, which can be attributed to its ability of reducing interference between different features. Although TCN can capture skip-correlated patterns to some extent, it requires relatively deep layers to conduct hierarchical convolutions to cover the entire sequence and convolution operations across different features over and over again may hurt the expression of the pattern of a particular feature; 3) Comparing SNN with DNN, the improvements of \textbf{10.5} Absolute Percentage (AP) on HR@3, and \textbf{9.8} AP on HR@10 suggest that modeling channel's pump history can bring a huge benefit to this task.
This performance boost can be easily extended to any other non-sequential methods, \textit{e.g.}, traditional ML models, by incorporating sequence representations extracted by a trained SNN.

\subsection{Coin Embedding Test}

Table~\ref{tab:q3} summarizes the results of six approaches on the testing set. E2E means a DNN model that only takes coin\_id as input, while CW and SG are the DNN models trained on CBoW and SkipGram embedding, respectively. SNN$_{C}$ and SNN$_{S}$ share the same architecture as SNN but the original coin\_id embeddings are replaced with CBoW and SkipGram embedding, respectively.

It is not surprising to see E2E gives the worst performance since it suffers from the cold-start problem. In comparison, both CW and SG show much superior performance, especially with SG achieving the HR@3 of 0.115, suggesting that well-trained word embedding can alleviate the cold-start problem.
This conclusion is also confirmed by the observation that SNN$_{C}$ and SNN$_{S}$ achieve significant positive lifts compared to the original SNN.

\subsection{Sequence Pattern Visualization}

We try to reveal some meaningful sequence patterns by visualizing the positional attention vectors of several features learned during training. Figure~\ref{fig:visualization}(a) presents the results of attention vectors for the target coin prediction task, where $F_1 \sim F_6$ are \textit{coin\_id}, \textit{volume}, \textit{price}, \textit{Twitter\_follower}, \textit{market\_cap}, and \textit{Alexa\_rank}, respectively. $P_1$ is the temporal closest position and $P_{30}$ is the farthest position. The darker the color, the higher the attention score. We observe that $F_1$ to $F_4$ exhibit skip-correlated patterns, which supports our design of positional attention module to capture this type of feature pattern. $F_5$ and $F_6$ exhibit the temporal proximity pattern, indicating that organizers tend to pump coins with similar market cap and web popularity that close to the coins pumped last few times.

\section{Generalizability of Methodology}

To assess the generalizability of our methodology, we extend the data science pipeline to a different task, namely sentiment-enhanced cryptocurrency price forecasting. 
In Telegram trading groups, we observe that investors share news and predict market movements. Their sentiment can be use to predict the market movement, or the market movement might be affected by their sentiment because their subsequent trading operations.


For this task, we use Bitcoin as an example, and the methodology can be easily extended to other coins. Following the same pipeline for target coin prediction, we explore several Telegram chat groups and collect user messages posted from Feb. 7, 2020, to May 31, 2022. Secondly, we apply a data filtering strategy to select Bitcoin-related messages. We extract sentiment information by VADER~\cite{vader} and calculate sentiment statistical features within each hour (\textit{avg\_score}, \textit{neg\_avg\_score}, \textit{neg\_num}, \textit{pos\_avg\_score}, \textit{pos\_num}, \textit{message\_num}, \textit{etc}). Finally, we construct a 200-hour-length sequence for each sample and use hourly price and sentiment features as the sequence feature. The label is Bitcoin's average price in the future 48 or 96 hours. The dataset description for this task is presented in Table~\ref{tab:BTC_dataset}.
We adopt Mean Absolute Error (MAE) as the object loss during training.
which is defined as:
\begin{equation}
L=-\frac{1}{|\mathcal{D}|} \sum_{(\hat{y}, y) \in \mathcal{D}} |y-\hat{y}|
\end{equation}
where $\mathcal{D}$ is the training set, $y$ is the ground-truth label and $\hat{y}$ is the prediction value.
In this task, SNN solely takes the sequence features as input for price forecasting.

\subsection{Experiment Settings}
\textbf{Dataset:} We collect Bitcoin's hourly price from Feb. 7, 2020, to May 31, 2022, which is the same period as the Telegram message data. We split the training and testing sets by the timestamp "2021-12-19 00:00:00". For each sample, we construct a 200-hour sequence with extracted sentiment features and hourly price. The label is the average price of Bitcoin in the future 48 or 96 hours, as predicting the price in the future 1 hour is considered too easy.

\noindent \textbf{Competitors:} We use the same sequential competitors as the target coin prediction task. For TCN, we set the depth of convolution layer to 5 with 16 channels per layer, the kernel size is set to 8 to cover a 200-length sequence. For SNN, we set the channel number to 16 for $hour\_price$, and for other features, the channel numbers are set to 2. For RNN-based methods, we set the hidden dimensions to 32 based on empirically hyper-parameter tuning.

\noindent \textbf{Evaluation Metrics:} We use MAE as the metric, which is the same as the objective loss optimized during training.

\begin{table}[tbp]
\centering
\caption{Statics of BTC Price Forecasting Dataset}
\setlength{\tabcolsep}{0.7mm}
\begin{tabular}{cccccc}
\toprule
\textbf{\# Message} & \textbf{\# BTC msg} &\textbf{\# Pos msg} & \textbf{\# Neg msg} & \textbf{\# Train} & \textbf{\# Test} \\ 
\midrule
2,799,669 & 229,595 & 88,512 & 54,175 & 15,856 & 3,964 \\
\bottomrule 
\vspace{0.1cm}
\end{tabular}
\label{tab:BTC_dataset}
\end{table}

\begin{table}[tbp]
\centering
\small
\caption{Performance Comparison for BTC Price Forecasting}
\setlength{\tabcolsep}{0.9mm}
\begin{tabular}{clp{0.15cm}p{0.15cm}p{0.15cm}p{0.15cm}p{0.15cm}p{0.15}}
\toprule 
\multicolumn{1}{l}{\textbf{Span}} & \multicolumn{1}{l}{\textbf{Metric}} & \multicolumn{1}{c}{\textbf{LSTM}} & \multicolumn{1}{c}{\textbf{BLSTM}} & \multicolumn{1}{c}{\textbf{GRU}} & \multicolumn{1}{c}{\textbf{BGRU}} & \multicolumn{1}{c}{\textbf{TCN}} & \multicolumn{1}{c}{\textbf{SNN}}     \\ \midrule
\multirow{3}{*}{48h} & MAE(P) & \multicolumn{1}{c}{871.21} & \multicolumn{1}{c}{810.87} & \multicolumn{1}{c}{851.30}& \multicolumn{1}{c}{812.45} & \multicolumn{1}{c}{820.32} & \multicolumn{1}{c}{\textbf{805.49}}  \\
& MAE(P+T)  & \multicolumn{1}{c}{848.29}  & \multicolumn{1}{c}{785.66} & \multicolumn{1}{c}{814.68} & \multicolumn{1}{c}{791.89} & \multicolumn{1}{c}{860.75} & \multicolumn{1}{c}{\textbf{756.90}}  \\
& Impr. & \multicolumn{1}{c}{22.92} & \multicolumn{1}{c}{25.21} & \multicolumn{1}{c}{36.62} & \multicolumn{1}{c}{20.56} & \multicolumn{1}{c}{-40.43} & \multicolumn{1}{c}{\textbf{48.59}} \\ \midrule
\multirow{3}{*}{96h} &  MAE(P) & \multicolumn{1}{c}{1144.23} & \multicolumn{1}{c}{1078.13} & \multicolumn{1}{c}{1126.37} & \multicolumn{1}{c}{\textbf{1049.85}} & \multicolumn{1}{c}{1059.36} & \multicolumn{1}{c}{1051.57} \\
& MAE(P+T) & \multicolumn{1}{c}{1118.84} & \multicolumn{1}{c}{1043.70} & \multicolumn{1}{c}{1088.25} &  \multicolumn{1}{c}{1027.45} & \multicolumn{1}{c}{1048.53} & \multicolumn{1}{c}{\textbf{964.27}}  \\
& Impr. & \multicolumn{1}{c}{25.39}  & \multicolumn{1}{c}{34.43} & \multicolumn{1}{c}{38.12} &  \multicolumn{1}{c}{28.40} & \multicolumn{1}{c}{10.83}  & \multicolumn{1}{c}{\textbf{87.30}}   \\ 
\midrule
- & Cost  & \multicolumn{1}{c}{4.68s}  & \multicolumn{1}{c}{5.41s} & \multicolumn{1}{c}{4.11s} & \multicolumn{1}{c}{4.61s} & \multicolumn{1}{c}{2.66s} & \multicolumn{1}{c}{\textbf{0.36s}} \\ 
\bottomrule
\label{tab:comparison_cpf}
\vspace{0.2cm}
\end{tabular}
\end{table}

\subsection{Performance Comparison}
\textbf{Performance Comparison:} Table~\ref{tab:comparison_cpf} summarizes the results on the testing set, where MAE(P) and MAE(P+T) indicate the MAE loss when taking only the hour price or with Telegram sentiment features as input, Impr. denotes the relative improvement, and Cost is the average training time for 50 batches. It can be observed that: 1) SNN outperforms all the other methods when incorporating Telegram features, suggesting that it maximally exploits the sentiment information; 2) SNN runs significantly faster as discussed in Section~\ref{sec:snn}; 3) The improvement brought by sentiment features for TCN is worse than other competitors, indicating that multiple times (5 layers to cover a 200-length sequence) of convolution operations might hurt the expressiveness of pattern of sentiment features.

\noindent \textbf{Sequence Pattern Visualization:} Figure~\ref{fig:visualization}(b) presents the visualization of positional attention for the price forecasting task, where $F_1 \sim F_6$ correspond to \textit{hour\_price}, \textit{neg\_num}, \textit{pos\_num}, \textit{message\_num}, \textit{neg\_avg\_score}, \textit{pos\_avg\_score}. We observe that $F_1$ strictly follows the temporal proximity pattern, and $F_2 \sim F_4$ show similar major patterns with slightly skip-correlated factors. On the other hand, $F_5$, $F_6$ demonstrate strong skip-correlation patterns, suggesting that the intensity of investor sentiment has a delayed impact on future price movement. Moreover, we find some of the attention channels of non-skip-correlated feature, \textit{e.g.}, \textit{hour\_price}, also demonstrates skip-correlated characteristics as $C_3$, $C_4$, $C_5$ presented in Figure~\ref{fig:visualization}(c). After manual review, it is find that some attention channels assign higher scores at the 24-th, 48-th positions, suggesting that through this way SNN captures the periodicity, a typical skip-correlation.


\section{Related Work}


\noindent \textbf{Anatomy:} Kamps et al.~\cite{kamps2018moon} are the first to introduce cryptocurrency P\&Ds and define their life cycle as accumulation, pump, and dump, and distinguish it from traditional P\&Ds in the stock market. 
Following their setting, Xu et al.~\cite{xu2019anatomy} present a comprehensive analysis of 412 P\&Ds that occurred during Jun. 2018 and Feb. 2019, focusing on describing how a P\&D activity is coordinated on Telegram.
Li et al.~\cite{li2021cryptocurrency} study P\&Ds from a statistical perspective and point out the detrimental influence of P\&Ds on the development of the cryptocurrency market.
Some characteristics of P\&Ds are also revealed and confirmed by the subsequent studies~\cite{hamrick2018economics,dhawan2020new,la2021doge}.
In addition, some works study P\&Ds from the perspective of social media. Nizzoli et al.~\cite{nizzoli2020charting} highlight the vast presence of Twitter bots and their key part in spreading the invitation links of pump channels in Telegram. Similar results revealed by~\cite{pacheco2021uncovering} show the widespread misinformation spread by Telegram channels to mainstream social media platforms.

\noindent \textbf{P\&D Post-detection:} This task aims to detect whether a P\&D has occurred or is ongoing, based on the statistics like price, trading volume, or signals from external sources, like social media. 
Kamps et al.~\cite{kamps2018moon} perform anomaly detection on the moving averages of price and trading volume data. La et al.~\cite{la2020pump,la2021doge} improve this method by introducing much more fine-grained features, like \textit{market buy orders}.
In addition to using market statistical data, Mirtaheri et al.~\cite{mirtaheri2021identifying} take into account the statistics of tweets and users that mentioned cryptocurrencies to make the detection. 
Even though some of the approaches~\cite{la2020pump,la2021doge} argue they can achieve real-time response, the post-detection paradigm fails to meet the practical needs since the coin price can easily reach its peak in less than one minute after a P\&D starts.

\noindent \textbf{Target Coin Prediction:} Xu et al. ~\cite{xu2019anatomy} are the first to introduce this task and they use a RF model to predict the target coin while treating P\&D activities as isolated, unrelated events. Nghiem et al.~\cite{nghiem2021detecting} focus on modeling the time-series market movement of target coin by BiLSTM and CNN algorithms, which differs from our fundamental setting.

\section{Conclusion}

In this paper, we present a novel data science pipeline to tackle the target coin prediction task. We identify 709 P\&Ds from millions of Telegram messages and extracts valuable and timely information from heterogeneous data sources to support efficient and effective target coin prediction. In our analysis, we discover some interesting phenomena, such as pumped coins exhibiting intra-channel homogeneity and inter-channel heterogeneity, which motivates us to propose SNN to encode a channel's pump history to accurately predict the target coin. Consequently, our method significantly improves over the base model, suggesting that the proposed data science pipeline is well-suited for real-world applications.

\section{Acknowledgement}

The authors would like to thank Lingling Lu, Jianting He and Yuan Chen from Zhejiang University for their valuable suggestions regarding this work.
This research was supported by the National Research Foundation, Singapore under its Industry Alignment Fund – Pre-positioning (IAF-PP) Funding Initiative. Any opinions, findings and conclusions or recommendations expressed in this material are those of the author(s) and do not reflect the views of National Research Foundation, Singapore.

\bibliographystyle{abbrv}
\bibliography{PD}

\end{document}